\documentclass[aps,prl,reprint,showpacs]{revtex4-1}
\usepackage{amsmath,amsfonts,amssymb,graphicx}

\newcommand{\beq}{\begin{equation}}
\newcommand{\eeq}{\end{equation}}
\newcommand{\bea}{\begin{eqnarray}}
\newcommand{\eea}{\end{eqnarray}}
\newcommand{\rem}[1]{ }
\newcommand{\bra}[1]{\left< #1\right|}
\newcommand{\ket}[1]{\left| #1\right>}
\newcommand{\V}{\mathcal{\hat V}}
\newcommand{\M}{\mathcal{\hat M}}
\newcommand{\Lb}{\mathbb{L}_\mathrm{box}}

\newcommand{\apjl}{Astrophys. J. Lett.}
\newcommand{\physrep}{Phys. Reports}

\begin{document}
\title{On the conversion of mass eigenstates}
\author{Mikhail V. \surname{Medvedev}}
\email{mvm@ias.edu}
\affiliation{Institute for Advanced Study, Princeton, NJ 08540}
\altaffiliation{Also at: Niels Bohr International Academy, Blegdamsvej 17, DK-2199 K\o benhavn \O, Denmark; Department of Physics and Astronomy, University of Kansas, Lawrence, KS 66045; Institute for Nuclear Fusion, RRC ``Kurchatov Institute'', Moscow 123182, Russia}

\begin{abstract}
In this paper we consider a stable particle with flavor mixing. We demonstrate that incoherent conversion of heavy mass eigenstates into light ones and vice versa can occur, as a result of elastic scattering. This effect is nontrivial for non-relativistic particles, for which the standard flavor oscillation ceases rapidly due to incoherence. We also prove that if a heavy state is bound in a gravitational potential and a light state is unbound, the mass-state conversion can lead to gradual ``evaporation'' of the mixed particle from the potential. A number of implications, ranging from the cosmic neutrino background distortions to scenarios of cold dark matter evaporation from halos, are addressed.
\end{abstract}
\pacs{03.65.-w, 14.60.Pq, 14.80.-j, 95.35.+d}
\maketitle

{\it Introduction} --- Massive flavor-mixed particles exist: the examples are quarks, $K^0$s, neutrinos, etc. Dark matter can also belong to this class. The propagation (mass) and interaction (flavor) eigenstates of mixed particles \cite{Pontecorvo58} are related by a unitary transformation,
\beq
\ket{f_i}=\sum_j U_{ij}\ket{m_j},
\eeq
where $\ket{f}$ and $\ket{m}$ denote the flavor and mass states, and $U$ is a unitary matrix. If the above superposition is coherent, then the time-dependent interference of mass states propagating with different speeds leads to the flavor oscillation phenomenon, which has been experimentally observed for relativistic neutrinos, for instance. In contrast to flavor eigenstates, mass eigenstates are more fundamental: they are thought to exhibit no  transformations (except for quantum broadening of their wave packets and/or decay into lighter particles if the primary is unstable) and are usually treated as ``real particles'' in a semi-classical sense. Indeed, the mass state wave packets have, in general, different energies and momenta and they propagate in the same way as the normal particles of masses $m_i$ would \cite{Nussinov76,*Kayser81}, whereas the observed pure flavor states are merely a time-dependent interference pattern. This view is also supported by the fact that it is impossible to construct a Fock space of flavor states, yet flavor oscillations can be derived without the concept of flavor eigenstates \cite{Giunti+92,*Giunti+93}. 

In this paper we show that transformations (or conversion) of mass states, $\ket{m_i}\to \sum_j a_j\ket{m_j}$ ($a_j$ being complex amplitudes), are possible even for stable particles. Repetitive conversions, schematically shown as
\beq
\begin{array}{llllllll}
\dots & \to & m_h & \to & m_h & \to & m_h & \to\dots \\
& &  & \searrow &  & \searrow & & \searrow \\
& & & & m_l & & m_l &
\end{array}
\label{evap}
\eeq
can have interesting implications for astrophysics and cosmology. We will make a number of assumptions throughout, which are useful for presentation purposes but are not crucial for the phenomenon at hand, hence they can be generalized or omitted.  

{\it Description} --- Let us consider a two-flavor system of stable particles, for simplicity. Without loss of generality, $m_1\ge m_2$. If $m_1\not= m_2$, we refer to them as the heavy and light states, $\ket{m_h}$ and $\ket{m_l}$, respectively. The interaction matrix, diagonal in the flavor basis, $\tilde V=\textrm{diag}(V_\alpha,V_\beta)$, is non-diagonal in the mass basis, $V=U^\dagger \tilde V U$. In particular, the off-diagonal terms are $V_{lh}=V_{hl}=(V_\alpha-V_\beta) \cos\theta\sin\theta$, where $\theta$ is the mixing angle and we assume $V_\alpha\not= V_\beta$. 

It is convenient to use the wave packet treatment, in which mass states are gaussians whose peaks follow classical trajectories. In general, mass states propagate with different velocities and can separate in the spatial domain. Such a mass state separation means that the wave packet overlap diminishes and the particle's wave-function becomes de-localized (not singly-peaked in space). Hence, the superposition of mass states is no longer coherent, so that the probability of detection of a certain flavor ceases to oscillate and tends to a constant, set by the mixing matrix. Along with the described ballistic mass-state separation, they can also separate via the (mass-dependent) interaction with gravitons or via their motion along different geodesics in a static gravity field (see more discussion below).

\begin{figure}
\includegraphics[angle = 0, width = 0.95\columnwidth]{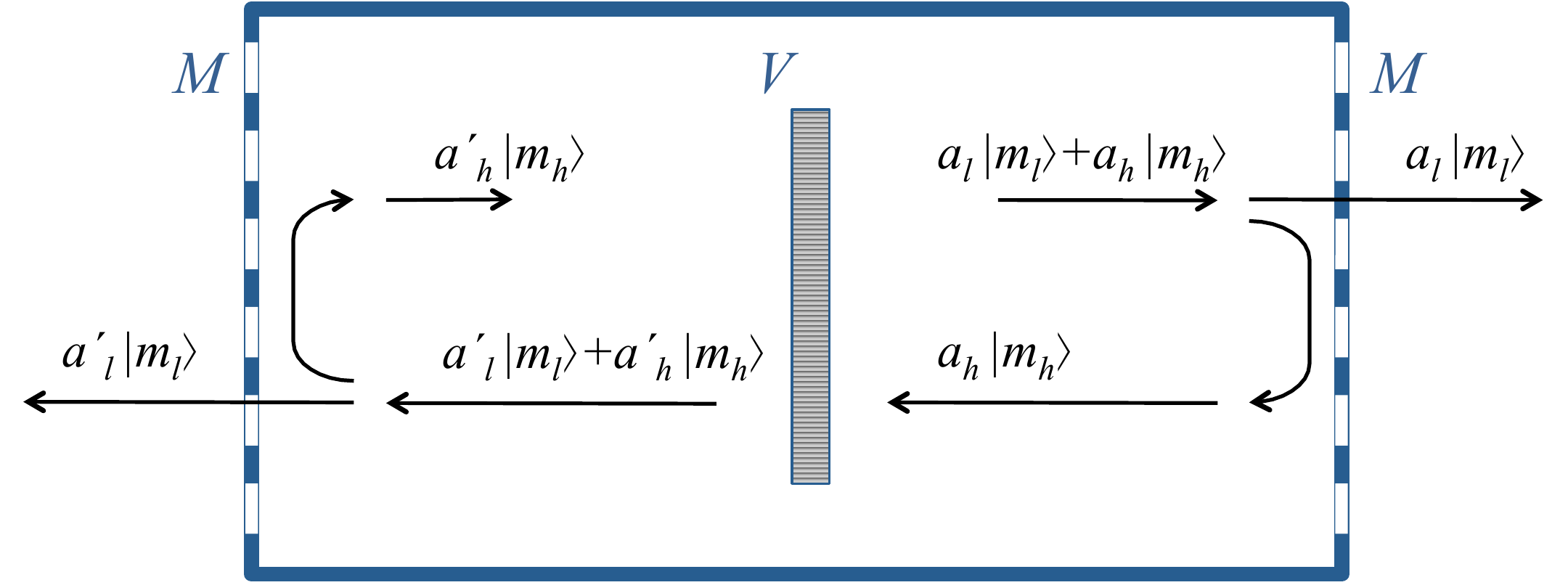}
\caption{Cartoon of a mixed particle in a box. Left and right ends of the box are ``membranes'', $M$, which keep a heavy state inside and let a light state to escape. A block in the center depicts a scattering potential $V$.}
\label{figbox}
\end{figure}

As an illustrative model, let us consider a particle in a one-dimensional box. The left and right ends of the box are ``membranes" (or semi-transparent ``mirrors'') that keep the heavy states inside but let the light states to freely escape from the box, see Fig. \ref{figbox}. Somewhere in the middle of the box, we put an elastic scatterer with an interaction matrix, $V$. We assume that the creation, detection and scattering processes are such that the widths of the mass state wave packets are always smaller than the box dimensions; this can be done by choosing a sufficiently large box. Now, consider a particle that bounces back an forth in the box and experiences elastic scattering from time to time. For generic initial conditions, the membranes let the light states escape from the box, so upon one transit time the box will contain a heavy state only. Because of off-diagonal terms in $V$, scattering of a single mass state yields a mixture of both states with smaller amplitudes. (Note that scattering shall not happen too frequently, because the mass states have to be well separated; otherwise the mass states may interact coherently, as a well defined flavor state.) Further evolution and interaction with a membrane leaves only a heavy state in the box again. Thus, by the end of the cycle, the system contains the same mass state but with a smaller amplitude. By repeating the cycle many times, one effectively converts the heavy mass state in the box into the light one represented by a ``wave-packet train'' outside the box, as in Eq. (\ref{evap}). In a sense, the particle ``evaporates'' from the box. We are not aware of any discussion of such a process in the literature.

This process can be described better using the operator formalism. The processes of elastic scattering and mass state separation due to propagation and interaction with a membrane  can be formally described by operators $\V$ and $\M$, which transform the mass state amplitudes and coordinates, respectively. Since the scattering potential is localized within the box, it has a compact support, $\mathrm{Supp}(V)\subset\Lb$, where $\mathbb{L}_\mathrm{box}=\{\mathbb{R}\ |\ 0\le x\le L_\mathrm{box}\}\subset\mathbb{R}$ is a one-dimensional compact manifold representing space inside the box of length $L_\mathrm{box}$. Therefore, we define the $\V$-operator as $\V\colon \{a_i\}\to\{a_i\}$ such that $\V\, \sum_i a_i(x_i)\ket{m_i}=\sum_i a_i'(x_i)\ket{m_i},\  i=l,h$ with $a_i'(x_i)=\hat V_{ij}a_j(x_j),\ x_i=x_j\ \forall x_j\in\Lb$ and $a'_i(x_i)=a_i(x_i)\ \forall x_i\not\in\Lb$, where the transition matrix elements $\hat V_{ij}\propto V_{ij}$ computed on the shell (``hat'' denotes transition amplitudes to distinguish them from the matrix elements of the scattering potential), $a_i(x_i)$ are complex-valued functions (representing shape factors and amplitudes of wave packets) of one argument $x_i$, respectively. The $\M$-operator is defined as $\M\colon \{a_i\}\to\{a_i\}$ such that $\M\,\sum_i a_i(x_i)\ket{m_i}=\sum_i a_i(x'_i)\ket{m_i},\ x_l,x_h,x'_h\in\Lb,\ x'_l\in\mathbb{R}\backslash\Lb$. Generalization to three dimensions and more than two flavors is straightforward. For the sake of simplicity, we also assume that the wave packets outside the box remain localized and never overlap with each other, as well as never come back into the box. We also simplify $\hat V_{ij}$ by assuming that one of the flavor states is non-interacting, $\hat V_\beta=0$, and another one is strongly interacting (with unit probability), $\hat V_\alpha=1$, and non-vanishing mixing is also assumed. Generalization of these assumptions is straightforward as well.

The following theorem holds true: 
Let $\Psi_k\equiv\sum_i a_i^{(k)}\ket{m_i},\ k=0,1$,  then $\forall a_i^{(0)}:\ \mathrm{Supp}(a_i^{(0)})\subseteq\Lb,\ \sum_i |a_i^{(0)}|^2=1,\ i=l,h,\ \exists a_i^{(1)}:\ \sum_i |a_i^{(1)}|^2=1$ that satisfy $\Psi_1=\M\V\M\,\Psi_0$, and the following is true: $|a_h^{(1)}|^2<|a_h^{(0)}|^2,\ |a_l^{(1)}|^2>|a_l^{(0)}|^2$ and $\mathrm{Supp}(a_h^{(1)})\subseteq\Lb$, where the norm is defined in a standard way: $|a_i|^2=\int a_i^* a_i dx_i$. That is, for an arbitrary initial state, $\Psi$, of a mixed particle which is initially in the box, the total detection probability of the heavy state decreases upon the action of the composite operator $\M\V\M$ upon $\Psi$, yet it remains identically zero outside the box. 

The proof follows from the direct substitution. First, let's label coordinates inside and outside the box as $x_i^\downarrow\in\Lb$ and $x_i^\uparrow\in\mathbb{R}\backslash\Lb$. Next, we have
\bea
\Psi_1&=&\M\V\M \sum_{i=l,h} a_i^{(0)}(x_i^\downarrow)\ket{m_i} \nonumber \\
&=&\M\V\left(a_l^{(0)}(x_l^\uparrow)\ket{m_l}+a_h^{(0)}(x_h^\downarrow)\ket{m_h}\right) \nonumber\\
&=&\M\left(a_l^{(0)}(x_l^\uparrow)\ket{m_l}+{a'}_l^{(0)}(x_l^\downarrow)\ket{m_l}+{a'}_h^{(0)}(x_h^\downarrow)\ket{m_h}\right) \nonumber\\
&=&\left(a_l^{(0)}(x_l^\uparrow)+{a'}_l^{(0)}(x_l^\uparrow)\right)\ket{m_l}+{a'}_h^{(0)}(x_h^\downarrow)\ket{m_h}.
\label{proof}
\eea
Thus, $a_h^{(1)}=a_h^{(0)}\hat V_{hh}$ and $a_l^{(1)}=a_l^{(0)}+a_h^{(0)}\hat V_{lh}$. Obviously, 
$|a_h^{(1)}|^2=|a_h^{(0)}|^2|\hat V_{hh}|^2<|a_h^{(0)}|^2$ for any nonzero mixing. Also, $|a_l^{(1)}|^2=|a_l^{(0)}|^2+|a_h^{(0)}|^2|\hat V_{lh}|^2>|a_l^{(0)}|^2$, where we used that wave packets do not overlap. Obviously,  $\sum_i |a_i^{(1)}|^2=|a_l^{(0)}|^2+|a_h^{(0)}|^2(|\hat V_{lh}|^2+|\hat V_{hh}|^2)=1$. Finally, from eq. (\ref{proof}), one has $\mathrm{Supp}(a_i^{(1)})=\mathrm{Supp}({a'}_i^{(0)})\subseteq\Lb$. q.e.d.

Clearly, multiple application of the $(\M\V)$-operator results in further reduction of the heavy state probability amplitude. If we define $(\M\V)^n\M=(\M\V)\dots(\M\V)\M$, then it is straightforward to show that for $\Psi_n=(\M\V)^n\M\,\Psi_0$ one has $|a_h^{(n)}|^2=|a_h^{(0)}|^2(|\hat V_{hh}|^2)^n\to0$ as $n\to\infty$. That is, a non-decaying  particle can be entirely converted into a light mass state and become unconfined. 

For completeness of the discussion, we qualitatively consider two more cases. Their quantitative analysis is beyond the scope of this paper. In case (A), let's assume the box is completely sealed (no membranes), so both mass states are kept inside. Then an equilibrium shall exist. There are two regimes: coherent and incoherent. If the wave packets are wide enough (their widths depend on the properties of production/detection, reflection and scattering processes), then they overlap and interfere coherently. Scattering is also coherent in this regime and the transition matrix elements shall be oscillatory functions of space and/or time. So, we expect that the equilibrium state shall depend on masses, energies and momenta of mass states, the size of the box and, perhaps, parameters of the scattering potential as well. In the second regime, when all wave packets are localized and their overlap is vanishing, the particle wave function is their incoherent sum. Then, scattering of mass states occurs independently, so an equilibrium is set by the detailed balance , $\ket{m_h}\rightleftarrows\ket{m_l}$. Equality of the rates of the forward and reverse processes, $a_h\hat V_{lh}=a_l\hat V_{hl}$, and the normalization condition $\sum_i |a_i|^2=1$ uniquely determine the equilibrium probabilities $|a_i|^2$, i.e., the mass and flavor composition of the system. In case (B), let's now put our original box with a particle (Fig. \ref{figbox}) into a larger sealed box of size $L_\mathrm{box,2}>L_\mathrm{box}$ and consider the incoherent regime only (note that light states can come back into the smaller box, but heavy states cannot go out into the larger one). The mass conversion is now skewed toward  $\ket{m_h}\to\ket{m_l}$ process, because the rate of the opposite process is reduced by a factor of $L_\mathrm{box}/L_\mathrm{box,2}$. Hence, the equilibrium composition is skewed towards the light states. In the extreme case of $L_\mathrm{box,2}\to\infty$, our original result, $|a_h^{(\infty)}|^2=0$, is recovered.

So far we considered an idealized model. In a realistic physical model one shall solve the Dirac equation to obtain accurate amplitudes and transition probabilities in the four-dimensional space-time. Note that in some quantum setups the $\V$ and $\M$ processes cannot be taken as independent. The analysis of such cases is interesting but is subject to a dedicated study elsewhere. We also want to keep the discussion fairly general, hence consideration of fine retails or specific models is beyond the scope of the paper. Here we consider a case when scattering and propagation are effectively decoupled, so the $\V$ and $\M$ processes are independent. 

The plane wave approximation is used for the $\V$. A free particle solution of the Dirac equation for $i$th mass state propagating in a certain direction is $\sim e^{iP_i X}$, which can be normalized by the box volume; here  $P=(\epsilon,\mathbf{p})$ and $X=(t,\mathbf{x})$ are four-vectors, and we use $\hbar=c=1$ throughout. In the perturbation theory, the matrix elements are $V_{ji}\sim\bra{m_j}\int d^4X\,e^{-iP'_jX}V({\bf x})e^{iP_iX}\ket{m_i}$, so that the time-dependent integral singles out into a $\delta$-function, $\int dt\,e^{i\epsilon'_jt-i\epsilon_it}\sim\delta(\epsilon'_j-\epsilon_i)$, which is just the conservation of energy. Note that $m_l\to m_h$ process can be kinematically suppressed if the primary ($m_l$) is slow, i.e., when $m_l^2+p_l^2<m_h^2$. In the $m_h\to m_l$ conversion, the secondary ($m_l$) has a larger momentum, $m_l^2+{p'}_l^2=m_h^2+p_h^2$. For a non-relativistic primary, the secondary can be relativistic and non-relativistic, respectively, 
\beq
p'_l\sim m_h \quad\textrm{and}\quad v'_l\sim (2\Delta m/m)^{1/2},
\label{v'}
\eeq
where $p_h$ was neglected, and in the latter case $m_h\approx m_l=m$ and $\Delta m=m_h-m_l$.

We now consider the physics of the $\M$-process. What in a real world would play a role of a box with membranes? We suggest to consider a gravitational potential of any origin (for instance, a star or a black hole, a galaxy with its dark matter halo, a cluster of galaxies, etc.). Obviously, one can always adjust the masses and momenta of $\ket{m_l}$ and $\ket{m_h}$, so that the heavy state is bound (trapped) in the potential ($v_h$ is smaller than the escape velocity) and the light state is unbound (e.g., relativistic or non-relativistic but just fast enough). Since the gravitation potential is macroscopic, the evolution can be treated in the wave-packet approximation \cite{PS00,*Giunti02} for each mass state. For simplicity, we consider a non-relativistic one-dimensional packet, 
\beq
\psi(x,t)=\exp\left\{i\eta_t(x-x_t)^2+ip_t(x-x_t)+i\phi_t\right\},
\eeq
where $\eta_t,\ \phi_t$ are complex, $p_t,\ x_t$ are real and all are functions of time. If $\langle\psi|\psi\rangle=1$, then, obviously, $x_t=\bra{\psi}x\ket{\psi}$ and $p_t=\bra{\psi}p\ket{\psi}$ take their simple classical meaning. In turn, $\eta_t$ determines the width of the wave packet and the phase $\phi_t$ can be shown to be the action along the path. For a locally quadratic potential, $U(x)=U_0+U'(x-x_t)+U''(x-x_t)^2$, i.e., with $U'''$ and higher terms being identically zero, where $U_0=U(x_t),\ U'=d_xU|_{x=x_t}$, etc., it can be shown by direct substitution into the Schr\"odinger equation $i\dot\psi=H\psi$ with the Hamiltonian $H=-(2m)^{-1}\partial_{xx}^2+U(x)$ that 
\bea
& &\dot x_t=\partial H/\partial p_t, \qquad \dot p_t=-\partial H/\partial x_t,
\label{a}\\
& &\dot\eta_t=-2\eta_t^2/m-U''/2, \quad \dot\phi_t=i\eta_t/m+\dot x_tp_t-E, 
\eea
where $E=H(x_t,p_t)=p_t^2/2m+U(x_t)$. Particularly important for us is the first pair, Eqs. (\ref{a}), which demonstrates classical motion of the packet. It is fair to say that mass states propagate along geodesics. Whether a mass state is trapped or not is determined by it's velocity and the escape velocity of the potential.

{\it Implications} --- Below we offer a few possible scenarios where mass conversions can occur. A more detailed analysis is beyond the scope of the paper.

The cosmic neutrino background (CNB) is the relic neutrinos \cite{KolbTurner,Dolgov08} with the present day temperature of $\lesssim2$~K. Although neutrino masses are unknown, one can estimate them from oscillations \cite{Bilenky+02} to be $m\sim(\Delta m^2)^{1/2}\sim0.05-0.007$~eV, implying that cosmological neutrinos are nonrelativistic and the heaviest ones can be trapped in dark matter halos with escape velocities of order a few thousand km/s. Hence, if one ever be able to detect CNB, one can notice that it's composition is distorted from the the expected equipartition between the flavors. Assuming the non-relativistic cosmological neutrino cross-section $\sigma_\nu\sim10^{-60}\textrm{ cm}^2$ (it can be greatly enhanced by coherent effects, however \cite{Dolgov08}), the mean density of nucleons in the halos $n\sim10^{-3}\textrm{ cm}^{-3}$, the thermal velocity of $v\sim10^{-2}$, we estimate the number of scatterings to be $\sim n\sigma_\nu v t_H\sim10^{-47}$ per particle in the Hubble time, $t_H=4\times10^{17}$~s. The effect is very tiny and unobservable (unless nonrelativistic neutrinos interact more strongly then we currently think), though it may have some effect for the ultimate fate of the Universe, at $t_H\to\infty$.

Another interesting possibility is axion-photon mixing, if axions constitute cold dark matter (CDM) \cite{KG10,*Bassan+10,KolbTurner}. In this scenario, axions can scatter off cosmic magnetic fields. Using an experimental upper limit on axion-photon coupling $g_{a\gamma}\sim10^{-11}\textrm{ GeV}^{-1}$, we estimate the interaction/conversion probability \cite{KG10,*Bassan+10} to be $\sim 0.1(g_{a\gamma}/10^{-11}\textrm{ GeV}^{-1})(B/1~\mu{\rm G})(L/10~{\rm kpc})$. For galactic halos with the typical galactic field strength of, say, $B\sim1-3~\mu{\rm G}$ and the halo sizes $L\sim30-100$~kpc, the probability is of order unity per passage. In galaxy clusters with $B\sim0.1~\mu$G and sizes of a couple of Mpc, the rate can be similar. Interestingly, this is the most optimal regime of conversions (one per passage) and the cosmological effect can be noticeable. If occurs, it can lead to substantial ``evaporation'' of axion CDM halos on the Hubble time-scale. If the effect is not observed, it can further constrain the coupling constant, $g_{a\gamma}$.

The third scenario deals with the CDM in the weakly interacting massive particle (WIMP) sector, which is still an attractive model \cite{A-H09,Feng06,KolbTurner}. Usually, it involves a stable lightest (tens of GeV--TeV-scale) supersymmetric particle -- supposedly a mixture of several supersymmetric particles, in fact. For high mass degeneracy, which we consider now, decay channels can be kinematically forbidden, so more than one particle can be stable. Taking rather crudely $\Delta m\sim\textrm{ MeV}$ and $m\sim\textrm{ TeV}$, we estimate the velocity of a light state after $m_h\to m_l$ conversion, Eq. (\ref{v'}), to be $v'_l\sim300\textrm{ km s}^{-1}$, which is a cosmologically interesting number. It is smaller or comparable to the escape velocity from halos of large galaxies and galaxy clusters, but it is larger then $v_{\bf esc}$ for halos of dwarf galaxies, which can be as low as a few tens km/s. A halo dichotomy is then expected: the number of small halos must be smaller than what CDM simulations predict whereas the population of large halos is mainly unaffected. Such an effect can be an alternative explanation to the `missing satellite problem' and, perhaps, it can also affect halo cores \cite{Kravtsov10,*KdN+10}. For the effect of conversions to be significant, the rate shall be at least a few reactions per Hubble time, which seem to rule out scattering off normal matter. Dark matter self-interaction can accommodate the needed large cross section, though simple models seem to be disfavored \cite{Kravtsov10,*KdN+10}. The models with the velocity-dependent cross-section and the Sommerfeld enhancement \cite{A-H09}, which is the most profound at low velocities (that is, in dwarfs, again), shall be tested before this scenario is ruled out entirely. The disadvantage of this scenario is also in that it required strong degeneracy, which needs an explanation. However, a similar but {\it ad hoc} model with GeV-scale WIMPs with a keV-scale mass-degeneracy seems to explain some direct detection experimental data \cite{Essig+10}.

\acknowledgments
The author thanks Nima Arkani-Hamed and Peter Goldreich for thoughtful discussions, Masataka Fukugita for interesting suggestions and Freeman Dyson for encouragement. 
The author acknowledges support from The Ambrose Monell Foundation (IAS) and The Ib Henriksen Foundation (NBIA), as well as grants AST-0708213 (NSF), NNX-08AL39G (NASA) and DE-FG02-07ER54940 (DOE).

%


\end{document}